\documentclass[12pt,amsmath,amssymb]{article}
\usepackage[utf8]{inputenc}
\usepackage[T1]{fontenc}
\usepackage{times}
\usepackage{amsmath}
\usepackage{amssymb}
\usepackage{amsfonts}
\usepackage{amsthm}
\usepackage{mathrsfs}
\usepackage{slashed}
\usepackage{hyperref}

\newcommand{\pd}[1]{\frac{\partial}{\partial #1}}

\newcommand{\vev}[1]{\langle #1 \rangle} 
\newcommand{\state}[1]{|#1\rangle}

\newcommand{\matel}[3]{\langle #1|#2|#3\rangle}
\newcommand{\VEC}[1]{\vec{#1}}
\newcommand{\SP}{\!\cdot\!}
\newcommand{\Lcc}{\Lambda_{\rm GT}}
\newcommand{\R}[1]{\bar{#1}}
\newcommand{\ph}{\varphi}
\newcommand{\Gauss}{{\cal G}}

\usepackage[all]{xy}
\begin{document}

\begin{titlepage}
\begin{flushright}\begin{tabular}{l}
Edinburgh/13/37 \\
CP$^3$-Origins-2013-051 DNRF90  \\ 
DIAS-2013-51
\end{tabular}
\end{flushright}

\vskip1.5cm
\begin{center}
  {\Large \bf \boldmath Gluon condensates from the Hamiltonian formalism} 
  \vskip1.3cm 
  {\sc Vladimir Prochazka$^{\,a}$\footnote{v.prochazka@sms.ed.ac.uk}  \&
    Roman Zwicky$^{\,b}$\footnote{roman.zwicky@ed.ac.uk}}
  \vskip0.5cm
  
  $^a$ {\sl Higgs centre for theoretical physics \\
  School of Physics and Astronomy, \\
  University of Edinburgh,   Edinburgh EH9 3JZ, Scotland} \\
 
  \vspace*{1.5mm}
\end{center}

\vskip0.6cm

\begin{abstract}
 We derive  recently obtained relations, relating  the logarithmic gauge coupling derivative of the hadron mass and  the cosmological constant to
  the matter and vacuum gluon condensates, within a Hamiltonian framework.  
  The key idea is a  canonical transformation which brings the relevant part of
  the Hamiltonian into a suitable form.  
  Furthermore we illustrate the relations within the Schwinger model 
  and ${\cal N}=2$ super Yang Mills theory (Seiberg-Witten theory).
  \end{abstract}

\end{titlepage}

\setcounter{footnote}{0}
\renewcommand{\thefootnote}{\arabic{footnote}}

\tableofcontents

\section{Introduction}

The Feynman-Hellmann theorem \cite{FHclan}, originally derived in quantum mechanics, applies
straightforwardly to quantum field theory in the case where the relevant part of the 
 Hamiltonian is known. 
 One such example is the fermion mass term of a gauge theory 
${\mathcal H}_m = m \bar q q $ e.g.  
\cite{Gasser:1979hf}. The Hamiltonian formalism of gauge theories is not straightforward because of the elimination of two degrees of freedom from 
the vector potential one of which is associated with the gauge freedom. 

In \cite{FHfun} a Feynman Hellman relation for the gauge coupling constant was obtained by combining the trace anomaly, renormalization group  equation (RGE)
and the Feynman Hellmann theorem for the fermion mass. The relations read \cite{FHfun}:  
\begin{alignat}{2}
\label{eq:FHfunM}
& \R{g} \pd{\R{g}} E_\ph^2  &=&  -\frac{1}{2}\matel{\ph}{\frac{1}{\R{g}^2}  \R{G}^2}{\ph}_c    \;,  \\[0.1cm]
\label{eq:FHfunL}
&    \R{g} \pd{\R{g}} \Lcc &=& -\frac{1}{2} \vev{\frac{1}{\R{g}^2}  \R{G}^2}_0
\end{alignat}
where $G^2 = G_{\mu \nu}G^{\mu \nu} $ is the field strength tensor squared, the subscript $c$ stands for
the connected part, $\ph$ denotes a physical state (normalisation to be specified below) and 
$\vev{X}_0 \equiv \matel{0}{X}{0}$ corresponds to  the vacuum expectation value throughout.  
The scheme dependence of the matrix elements 
on the right hand side is determined by the scheme dependence of 
the couplings on the left hand side.
The barred symbols denote \emph{renormalized} quantities to distinguish them from unrenormalized  quantities. The partial derivatives are understood in the sense of the RGE. 
That is to say  implicit dependencies of other parameters on the coupling 
are not considered by definition. 
In Eq.~\eqref{eq:FHfunM} 
the momentum is taken to be independent of $M_\ph$ as in \cite{FHfun}.\footnote{The latter is of significance (appendix \ref{app:trace_ham}) for the derivation of
the trace anomaly matrix element from an RGE for the Energy.} 
  Relation \eqref{eq:FHfunM} is valid 
for the  following normalisation of states,
 \begin{equation}
  \label{eq:normal}
  \vev{\ph(E',\vec{p'})|\ph(E,\vec{p})} = 2 E_\ph   (2 \pi)^{D-1} \delta^{(D-1)}(\vec{p}-\vec{p'}) \;,
\end{equation}
where $D$ stands for the space-time dimension.  The cosmological constant $\Lcc$ 
 contribution in \eqref{eq:FHfunL}  was defined as $\vev{T_{\phantom{x}\mu}^{\mu}}_0 = D\Lcc$.  The goal of this paper is to derive these relations, after all, using a Hamiltonian formalism. The key observation is that by a  canonical transformation (rescalings in the gauge coupling constant), 
one can obtain a suitable form of the Hamiltonian.

The paper is organised as follows. In section \ref{sec:Hamilton} we pursue 
the derivation of relations (\ref{eq:FHfunM},\ref{eq:FHfunL}) within the Hamiltonian formalism. In sections \ref{sec:QED2}, \ref{sec:QED2mm} and \ref{sec:SW} we illustrate the formula within the Schwinger Model and the ${\cal N}=2$ super Yang Mills theory (Seiberg-Witten theory).  We end the paper with  
summary and conclusions in section 
\ref{sec:con}.
Relevant comments on the transformation of the measure under the canonical transformation can be found in appendix
\ref{app:Konishi}.

\section{(Re)derivation in the Hamiltonian formalism}
\label{sec:Hamilton}

\subsection{The suitable canonical transformation of the Hamiltonian}

In the Hamiltonian formalism of a (non-abelian) gauge theory $\VEC{\pi} = \VEC{E}$  and
$\VEC{A}$ 
are the independent canonically conjugate variables. 
 (e.g. \cite{Ramond:1981pw}).\footnote{The variable $A_0$ is degraded to be a 
 Lagrangian multiplier imposing Gauss' law in 
 (c.f.  ${\mathcal H}_\Gauss$ below)
 and $\pi_0 = 0$ is at the heart of 
 all the difficulties with the Hamiltonian formalism of gauge theories (parameterised by ${\mathcal H}_C$ below).} 
  The Hamiltonian reads,
  \begin{alignat}{2} 
\label{Ham}
& \mathcal{H} &\;=\;&  \mathcal{H}_g + 
\mathcal{H}_{C} +   \mathcal{H}_{\Gauss}  \;, \nonumber \\[0.1cm]
&\mathcal{H}_g &\;=\;& \frac{1}{2}(\VEC{E}^2 + \VEC{B}^2)  -  
 \overline{ q}( i \VEC{\gamma} \SP  \VEC{D}  -m) q   \;,
\end{alignat}
where $\VEC{D}  = \VEC{\partial} + i g \VEC{A}$ is the gauge covariant derivative 
and $q$ stands for fermions (quarks) in some representation of the gauge group. 
The magnetic field is defined as $2 B_k = \epsilon_{kij} G_{ij} =  \epsilon_{kij} ( \partial_i A_j - \partial_j A_i + i g [A_i,A_j])$. The term 
$\mathcal{H}_{\Gauss} = A^a_{0} \Gauss^a$ with 
$\Gauss^a = ( (\VEC{D} \SP \VEC{E})^a + 
\bar q t^a  \gamma_0 q)$ corresponds to Gauss' law 
 (i.e. one of Maxwell's equations).
The expression 
 $\mathcal{H}_{C}$ is  associated with primary and secondary constraints (resulting in gauge transformation). Both $\mathcal{H}_{\Gauss}$ and 
 $\mathcal{H}_{C}$ vanish on matrix elements of physical states 
 and shall therefore be omitted hereafter.
 
   Our strategy is to make the dependence on the coupling $g$ as simple as possible 
 through the canonical transformation,
 \begin{align} \label{eq:canonical} 
\VEC{A} &\rightarrow \frac{1}{g} \VEC{A}  \nonumber \\
\VEC{E} &\rightarrow g \VEC{E}  \;.
\end{align}
This leads to a Hamiltonian of the form,
\begin{equation} \label{eq:dgH}
\mathcal{H}_g= \frac{1}{2}(g^2\VEC{E}^2 + \frac{1}{g^2} \VEC{B}^2) -  
 \overline{ q}(  i  \VEC{\gamma} \SP\VEC{D}  +m) q      \;,
\end{equation}
where, crucially, the only $g$-dependence is in front of the electric and magnetic field terms.
 It is important to note that the transformation in Eq.~\eqref{eq:canonical} 
leaves the measure of the path integral ${\cal D} \vec E {\cal D} \vec A$ invariant.
First the transformation \eqref{eq:canonical}  does not affect the equal time canonical commutation relation, 
$[ A^k(x_0,\VEC{x}) ,E_l(x_0,\VEC{y})] = i \delta^k_{\phantom{l}l} \delta^{(D-1)}(\VEC{x} - \VEC{y})$; 
the (simple) Jacobian is therefore trivial.
Second the measure is not affected by a rescaling anomaly 
of the type \cite{Konishi} since the two transformations in  \eqref{eq:canonical}
exactly cancel each other (as outlined in appendix \ref{app:Konishi}).
\subsection{Gluon condensates from Hamiltonian}
The Feynman-Hellmann theorem \cite{FHclan} in quantum mechanics (here $\vev{\ph|\ph} = 1$) 
states  that 
\begin{equation}
\label{eq:FH}
\pd{\lambda} E_\ph(\lambda) = \matel{\ph}{\pd{\lambda} H(\lambda)}{\ph} \;,
\end{equation}
where $\lambda$ is a parameter.
It is crucial that $\state{\ph}$ is an eigenstate of the Hamiltonian $H$.  
The rest follows from  the 
normalisation being independent on the parameter $\lambda$. 
The adaption to quantum field theory solely involves  the incorporation  of the specific normalisation 
convention   \eqref{eq:normal}.
The right hand side of \eqref{eq:FH}, in our case, is obtained by 
 differentiating \eqref{eq:dgH}\begin{equation}
\label{eq:almost}
g \pd{g} \mathcal{H}_g =  g^2\VEC{E}^2 - \frac{1}{g^2} \VEC{B}^2 = 
-\frac{1}{2}\frac{1}{g^2} G_{\mu \nu} G^{\mu \nu} \;.
\end{equation}
This form is very close to  Eqs.~(\ref{eq:FHfunM},\ref{eq:FHfunL}). 
In particular a Lorentz invariant result has emerged from the non-covariant 
Hamilton formalism as is usually the case. 
Note, the Hamiltonian is a physical quantity 
and is therefore not renormalized. 
Below we shall write the Hamiltonian in terms of renormalized quantities (denoted by bars) which is natural since the physical quantities are matrix elements thereof.
Identifying $\vev{\mathcal{H}}_0 = \Lcc$ one gets \eqref{eq:FHfunL} 
from \eqref{eq:almost}:
\begin{equation}
\R{g} \pd{\R{g}} \Lcc  = \vev{\R{g} \pd{\R{g}} \mathcal{H}}_0 +\Lcc \, \R{g}  \underbrace{ \pd{\R{g}} \vev{0|0}}_{=0} 
\stackrel{\eqref{eq:almost}}{=}  -\frac{1}{2} \vev{\frac{1}{\R{g}^2}  \R{G}^2}_0 \;.
\end{equation}
For the derivation of \eqref{eq:FHfunM} the factor $E_\ph$ in the normalisation \eqref{eq:normal} complicates the algebra and we shall use $\sqrt{2 E_\ph}\state{ \tilde \ph } = \state{\ph}$ below restoring the factor in the end.  
\begin{equation*}
\label{eq:Dnormal}
 \R{g} \pd{\R{g}} E_\ph = 
 \R{g} \pd{\R{g}}    \matel{\tilde \ph}{\mathcal{H}}{ \tilde \ph}_c = 
\matel{\tilde \ph}{\R{g} \pd{\R{g}}  \mathcal{H}}{\tilde \ph}_c + 
 \frac{E_\ph}{V} \, \R{g}  \underbrace{  \pd{\R{g}} \vev{\tilde \ph|\tilde \ph}_c}_{= 0 }  =  \matel{\tilde \ph}{\frac{1}{\R{g}^2}  \R{G}^2}{\tilde \ph}_c
\end{equation*}
where $V$ is the volume. Above we have identified $(2 \pi)^{D-1} \delta^{(D-1)}(\vec{p}-\vec{p'})  = \int d^{D-1} x $ (in the sense of distributions) since the Hamiltonian is given by $H= \int d^{D-1} x \mathcal{H}$. Restoring the normalisation \eqref{eq:normal} we get an expression,
\begin{equation}
 2 E_\ph  \R{g} \pd{\R{g}} E_\ph =  \matel{ \ph}{\frac{1}{\R{g}^2}  \R{G}^2}{ \ph}_c
  \;,
\end{equation}
which is equivalent to \eqref{eq:FHfunM}.  We have therefore rederived 
Eqs.~(\ref{eq:FHfunM},\ref{eq:FHfunL}) in a Hamiltonian framework which was 
the main goal of our work. We proceed to illustrate the formula in three models 
where exact results are known.

\section{Examples}

The relation \eqref{eq:FHfunM} was used \cite{size}
 to derive  the scaling corrections to the hadron masses 
  in two  alternative ways. It therefore constitutes  one independent check.
Below we provide three further examples.

\subsection{Photon mass in the Schwinger Model}
\label{sec:QED2}

Two dimensional quantum electrodynamics, known as 
the Schwinger model \cite{SW62,SW63} (for a review c.f. \cite{AAR-book}), has served as a test ground 
for many formal approaches and lattice simulations.
A curious feature of the Schwinger model is that the photon acquires a mass through 
the chiral anomaly as the $\eta'$ in quantum chromodynamics. 
This is sometimes referred to as  a dynamical Higgs mechanism. The photon mass is:
\begin{equation}
\label{eq:MS}
M_\gamma^2 = \frac{e^2}{\pi} \;.
\end{equation}
The relation \eqref{eq:FHfunM} adapted to the Schwinger model, for a massive photon state at rest,  reads:
\begin{equation}
\label{eq:GMS}
e\pd{e} M_\gamma^2 = -\frac{1}{2} \matel{\gamma}{F^2}{\gamma}_c \;.
\end{equation}
Above $F^2 = F_{\mu \nu} F^{\mu \nu} $  is the electromagnetic field strength tensor squared and $e$ is the charge of mass dimension one. The latter does not receive any renormalization (vanishing beta function).  

In order to obtain \eqref{eq:MS} from \eqref{eq:GMS} we have to evaluate the matrix element 
$ \matel{\gamma}{F^2}{\gamma}_c$ for which we resort 
to the operator solution of the Schwinger model \cite{LS71} (e.g. chapter 10 \cite{AAR-book}).  The Field strength tensor is given by
\begin{equation}
\label{eq:F}
F_{\mu \nu} = \frac{\sqrt{\pi}}{e} \epsilon_{\mu\nu} \Box  \Sigma \;,
\end{equation}
where $\Box = \partial_\mu \partial^\mu$ is the Laplacian 
and $\Sigma$ is a canonically normalised free field of mass $e^2/\pi$.   
Choosing the connected part automatically fixes the scheme of the matrix element,
which incidentally  corresponds to normal ordering as used in ordinary perturbation theory: 
$ \vev{F^2}_0 = 0$. 
This is not surprising since there is no scheme ambiguity on the left hand side as 
the coupling does not run.
Through an explicit computation in terms of creation and annihilation operators one gets,
\begin{equation}
\label{eq:M-F2}
\matel{\gamma}{F^2}{\gamma}_c = \frac{\pi^2}{e^2} \epsilon_{\mu \nu} \epsilon^{\mu \nu} 2(- M_\gamma^2)^2 = -4 \frac{e^2}{\pi} \;,
\end{equation}
where the factor of $2$ is of combinatorial nature and we have replaced $\Box \to 
- q^2 = - M_\gamma^2$. 
Inserting \eqref{eq:M-F2} into \eqref{eq:GMS} we get:
\begin{equation}
\label{eq:again}
e\pd{e} M_\gamma^2 = 2 \frac{e^2}{\pi}  \quad \Rightarrow \quad  M_\gamma^2 = \frac{e^2}{\pi} + C \;,
\end{equation}
where $C$ is a constant. From the limit $ e\to 0$, where we expect $M_\gamma \to 0$, we infer $C=0$ and therefore  \eqref{eq:again}
corresponds to the exact result \eqref{eq:MS} known in the literature. 
In essence we have shown that \eqref{eq:F} and \eqref{eq:GMS}  implies the Photon 
mass \eqref{eq:MS}.

As an additional, but not necessary test, we can verify whether \eqref{eq:GMS} is compatible with an RGE.  
The trace of the energy momentum tensor in massless
QED, in terms of bare quantities,  reads $T^\mu_{\phantom{x} \mu} = -(D-4) {\cal L} + \text{EOM}$, where 
$\text{EOM}$  stands for terms which vanish by the equation of motions. 
The latter are not of interest for us as we shall evaluate the trace on physical states. Using $D= 2$ we get
\begin{equation}
\label{eq:extend}
 \matel{\gamma}{T^\mu_{\phantom{x} \mu}}{\gamma}_c = -\frac{1}{2} \matel{\gamma}{F^2}{\gamma}_c \;,
\end{equation}
and since $ 2 M_\gamma^2 =  \matel{\gamma}{T^\mu_{\phantom{x} \mu}}{\gamma}_c$ it can be combined with \eqref{eq:GMS} into
\begin{equation}
\label{eq:RGEdim}
(e\pd{e}  -2 ) M_\gamma^2  = 0 \quad \Rightarrow \quad 
 M_\gamma^2 = 
C' \, e^2 
\end{equation}
where $C'$ is a constant ($C' = 1/\pi$ according to \eqref{eq:MS}) and the equation 
on the right hand side corresponds to an RGE. In fact the latter is equivalent to 
an equation based on dimensional analysis on grounds of the fact that there are 
no running quantities in the Schwinger model. 

\subsection{Vacuum energy in massive mutliflavour Schwinger model}
\label{sec:QED2mm}

The Schwinger Model with $N_f$ massive fermions has aspects which are known exactly (c.f. \cite{smilga} and references therein).
The model has got a global  $SU_L(N_f) \times SU_R(N_f)$ flavour symmetry which is explicitly broken 
down to $SU_V(N_F)$
 by the fermion mass term. The spectrum consists of one massive boson (the massive 
photon of the proceeding section) and $N_f^2-1$ quasi Goldstone boson, similar to the $\eta'$ and the octet $\pi,K,\eta$ in quantum chromodynamics.  The situation is though distinct 
in that the quark condensate does not form in the massless case and the quasi Goldsone bosons
show scaling behaviour of a critical theory. The vacuum energy is proportional to the mass gap squared (for $m \ll e$ c.f. \cite{smilga} and references therein):
\begin{equation}
\Lcc \propto M^2_{\rm gap} \propto m^{\eta_m} e ^{\eta_e} \;, \quad  
\eta_ m = \frac{2 N_f}{N_f +1} \;, \;\; \eta_e =\frac{2 }{N_f +1} \;. 
\end{equation}
From the trace anomaly equation one gets:
\begin{eqnarray}
2 \Lcc = -\frac{1}{2} \vev{F^2}_0 + N_f m \vev{\bar q q}_0 \;.
\end{eqnarray}
The analogous equation for  four dimension is given in \cite{FHfun}. 
The adaption of the $F^2$-term to two dimensions has been discussed in the previous section 
and the anomalous dimension of the mass is zero. 
Using \eqref{eq:FHfunL} and $N_f m_f \vev{\bar q q }_0 = m \pd{m} \Lcc$ one gets
\begin{eqnarray}
\label{eq:entlarvt}
2 \Lcc =  e \pd{e} \Lcc + m \pd{m}  \Lcc 
 = (\underbrace{ \eta_e  + \eta_m}_{=2}) \Lcc \;,
\end{eqnarray}
a consistent result.
Summarising we obtain $\vev{F^2}_0 = -2 \eta_e \Lcc$ and $N_f m \vev{\bar q q}_0 =  \eta_m \Lcc$.
Again \eqref{eq:entlarvt} reveals itself directly equivalent to an RGE for $\Lcc = \Lcc(m,e)$
\begin{equation}
\label{eq:RGEagain}
\left( e\pd{e} + m\pd{m} - \Delta_{\Lcc} \right)  \Lcc(m,e) = 0 \;.
\end{equation}
Above $ \Delta_{\Lcc} = 2$ is the scaling dimension of the $\Lcc$ which is free from anomalous scaling as it is an observable. As \eqref{eq:RGEdim} Eq.~\eqref{eq:RGEagain} is merely an equation that follows 
from dimensional analysis since all the scale breaking is explicit and not anomalous. 

\subsection{Magnetic monopole  in Seiberg-Witten theory}
\label{sec:SW}

The ${\cal N}=2$ pure super Yang-Mills theory (with gauge group $SU(2)$), known as Seiberg-Witten theory \cite{SW}, has features which are known exactly.  In particular it is known
that BPS states obey \cite{SW},
\begin{equation}
\label{eq:BPS}
M_{(n_e,n_m)} = 2 |Z|^2 \quad  \text{with}  \quad Z = n_e a + n_m a_D \;,
\end{equation}
where $n_e$ and $n_m$ count the units of electric and magnetic charges.
Exact solutions for $a$ and $a_D$ along with the effective coupling constant $\tau(a)$ constitute part of the work of Seiberg and Witten \cite{SW}. 
First we are going to derive Eq.~\eqref{eq:almost}  for the BPS sector. 
In the magnetic BPS sector the relevant part of the Hamiltonian  reads \cite{SW} 
\begin{equation}
\label{eq:HBPS}
\mathcal{H}_{\rm BPS} = \frac{1}{g^2} \VEC{D}\phi \SP \VEC{D}\phi + \frac{1}{2}\frac{1}{g^2} \VEC{B}^2 \;,
\end{equation}
where we shall comment on the (non-)significance of the extra $1/g^2$-factor 
in front of the scalar kinetic term shortly below. 
Note, Maxwell's equations imply $\VEC{E}=0$  for  static solution with $\VEC{B} \neq 0$ (magnetic monopole). 
The fermionic terms are absent by construction of what is known as a BPS state 
in supersymmetry.  
Using the BPS equation,
\begin{equation}
\label{eq:BPScond}
\vec{D} \phi \state{{\rm BPS}} = \frac{1}{\sqrt{2}} \VEC{B} \state{{\rm BPS}} \;,
\end{equation}
the total Hamiltonian becomes,
\begin{equation}
\mathcal{H}_{\rm BPS} = \frac{1}{g^2} \VEC{B}^2 \;,
\end{equation}
and the $\mathcal{N}=2$ supersymmetry, which is responsible for the $1/g^2$-factor 
in front of the kinetic term in \eqref{eq:HBPS},  effectively introduces a factor of $2$ 
in the relation \eqref{eq:FHfunM}.
This can be seen explicitly by differentiating, with respect to the coupling constant \eqref{eq:almost},
\begin{equation}
\label{eq:SWalmost}
g \pd{g} \mathcal{H}_{\rm BPS} 
 =  -2 \frac{1}{g^2} \VEC{B}^2  \stackrel{\VEC{E}=0}{=} -  \frac{1}{g^2} G^2
\end{equation}
and comparing with Eq.~\eqref{eq:SWalmost}.
In summary we have shown that in Seiberg-Witten theory \eqref{eq:FHfunM} holds
on the BPS subspace. Conversely assuming that the formula \eqref{eq:FHfunM} is true
we know that \eqref{eq:BPScond} has to hold for $\mathcal{H}_{\rm BPS} $ in \eqref{eq:HBPS}.

Unlike in the Schwinger model we cannot compute the matrix elements in 
\eqref{eq:SWalmost}  on the BPS states directly. 
We may turn things around and use the formula to express the matrix elements 
for the magnetic monopole in terms of $a_D$ which is known  explicitly in terms
of the coupling constant. 
Formula \eqref{eq:FHfunM} adapted for $\mathcal{N}=2$ supersymmetry (with factor of two difference as explained above) reads:
\begin{equation} 
\label{eq:BPSformula}
\matel{(0,n_m)}{\frac{1}{g^2} G^2}{(0,n_m)}_c = -  g  \frac{\partial  M^2_{(0,n_m)}}{ \partial g}
   \;,
\end{equation}
In order to evaluate the right hand side we use
 $M^2_{(0,n_m)} = 2 n_m^2 |a_D|^2$  \eqref{eq:BPS}  
 and  $g \frac{ \partial}{\partial g}= -\frac{1}{2} \omega \frac{\partial}{\partial \omega}$ 
 where $\omega \equiv \frac{1}{g^2},\,$\footnote{In doing so use the fact that  $a_{D}$ is a holomorphic function of holomorphicity in $\tau = 4 \pi i/g^2 + \frac{1}{2 \pi}  \theta$.}
\begin{eqnarray}
\frac{1}{n_m^2} \frac{\partial  M^2_{(0,n_m)}}{ \partial \omega}&=& 2[ a_D^{*} \frac{\partial a_D }{ \partial \omega}+a_D \frac{\partial a_D^{*} }{ \partial \omega}]=
8 \pi i [a_D^{*} \frac{\partial a_D}{ \partial \tau} - a_D \frac{\partial a_D^{*}}{\partial \tau^{*}}] \nonumber  \\
&=& - 16 \pi  \operatorname{Im}  [a_D^{*} \frac{\partial a_D}{ \partial \tau}] \;.
\end{eqnarray}
This leads to 
\begin{equation}
\matel{(0,n_m)}{\frac{1}{g^2} G^2}{(0,n_m)}_c= 8 \pi  \frac{n_m^2}{g^2} \operatorname{Im}  [a_D^{*} \frac{\partial a_D}{ \partial \tau}] \;.
\end{equation}
 The function $a_D$ is known \cite{SW}
\begin{equation}
 a_{D}(\tau)= \frac{\sqrt{2} \Lambda}{\pi}\int_{1}^{v(\tau)} \frac{dx \sqrt{x-v(\tau)}}{\sqrt{x^2-1}} 
 \;, \quad v(\tau) = -1 +  \frac{2}{\lambda(\tau)} \;,
\end{equation}
with $ v= u/\Lambda^2$ where $u = \vev{\phi^2}_0$ is a modulus and $\Lambda$ is a dynamical scale and constitute important parameters of the theory. The function 
$\lambda(\tau)$ is given in \cite{Brand}. 
We have checked numerically that the condensate is zero for $g_D \propto 1/g \to 0$ and increases monotonically as a function of $g_D$. The coupling $g_D$  corresponds to  the magnetic coupling and is dual to the electric coupling $g$. 
Loosely speaking the magnetic monopole condensate 
is governed  by the magnetic coupling $g_D$.  \\

\section{Summary and conclusions}
\label{sec:con}

We have derived the relations in Eqs.~(\ref{eq:FHfunM},\ref{eq:FHfunL}), previously obtained 
in \cite{FHfun} through the trace anomaly, the Feynman-Hellmann theorem and an RGE, 
in a Hamiltonian formulation of gauge theories.  
The derivation contains two ingredients. 
Eliminate the terms which 
vanish as matrix elements from the Hamiltonian. 
In this way we bypass the notoriously difficult problem of gauge fixing. 
The second step is a canonical transformation which arranges the Hamiltonian in such a way that only 
the $\VEC{E}^2$ and $\VEC{B}^2$-terms depend on the gauge coupling. The derivative with respect 
to the gauge coupling then gives rise to the explicitly Lorentz invariant result. 
A subtle point, which we have verified in appendix \ref{app:Konishi}, is that the canonical transformation 
is free from rescaling anomalies of the Konishi type. 
One possible advantage of the Hamiltonian 
derivation is that it makes it clear that the relations holds for gauge theories with more than one 
gauge coupling.
Furthermore  we have tested the relation within the Schwinger Model and the ${\cal N}=2$ super Yang Mills theory (Seiberg-Witten theory).  \\

{\bf Acknowledgements:} We are grateful to Arjun Berera, Luigi Del Debbio, Stephan D\"urr  and Donal O'Connell for useful 
discussions. 
R.Z. is grateful for partial support through an advanced STFC fellowship.

\appendix
\setcounter{equation}{0}
\renewcommand{\theequation}{A.\arabic{equation}}
\section{The rescaling anomaly in Hamiltonian language}
\label{app:Konishi}

In section \ref{sec:Hamilton} we have used a particular canonical transformation 
\eqref{eq:canonical} and one might wonder whether the measure is anomalous 
under this transformation. Generally any rescaling of a field which is gauged, 
produces anomalous term proportional to the kinetic term of the corresponding 
gauge field \cite{Konishi:1985tu}.  We shall see that for the transformation \eqref{eq:canonical} the effect cancels. 

Let us write \eqref{eq:canonical} for a generic transformation
 \begin{align} \label{rescaling1} 
\VEC{A} &\rightarrow \frac{1}{f(g)} \VEC{A}  \nonumber \;, \\
\VEC{E} &\rightarrow f(g) \VEC{E}  \;.
\end{align}
The anomalous Jacobian of the ${\cal D} \VEC{E} {\cal D} \VEC{A}$ measure is 
 \begin{eqnarray} \label{jacobian}
& &  \ln \det \frac{\delta Q'(x)}{\delta{Q(y)}} = 
 \ln \det \left( \begin{matrix} f(g)^{-1} \delta(x-y) &   0 \\ 0 & f(g) \delta(x-y) \end{matrix} \right) =
\nonumber \\ 
& &    \ln \det  \left( \begin{matrix} f(g)^{-1} &   0 \\ 0 & f(g) \end{matrix} \right) \delta(x-y) = \ln \det \delta(x-y) \;,
\end{eqnarray}
where we have used the compact notation $ Q \equiv (\VEC{A},\VEC{E}$).
It is proportional to an expression independent of $f(g)$ and therefore justifies
our manipulations in section \ref{sec:Hamilton}.
The  second equality sign is the crucial step where we use 
the fact that the $\VEC{A}$ and $\VEC{E}$ can be expanded in 
the same set of eigenfunctions.  For the chiral anomaly this is not the case since left and right handed fermions have different eigenfunction, or more precisely a different 
number of zero modes. For an arbitrary rescaling the two dimensional matrix 
on the second line is not of unit determinant and will therefore depend on the 
transformation \cite{Konishi:1985tu}.

\section{Trace anomaly  and the Hamiltonian}
\label{app:trace_ham}

In this appendix we show how the matrix element of the trace anomaly 
follows from an RGE of the Hamiltonian matrix elements. 
We consider  
\begin{equation}
\label{eq:h}
h(g,m,\mu,p) \equiv \matel{\ph(p)}{\mathcal{H}}{\ph(p)}_c  \stackrel{\eqref{eq:normal}}{=} 2 (E_\ph(p))^2 \;,
\end{equation}
where $p = |\vec{p}|$ denotes the spatial angular momentum which 
is considered to be an external parameter. By the latter we mean that it is 
in particular \emph{independent}
on $M_\ph$  in accordance with the remark below Eq.~\eqref{eq:FHfunM}. 
This type of matrix element satisfies an RGE of the form (e.g. \cite{RGE-Weinberg})
\begin{equation}
\label{eq:1}
( \R{\beta} \pd{\R{g}} - \R{m}(1+ \R{\gamma}) \pd{\R{m}} + \Delta_h - p \pd{p} ) 
h(\R{g},\R{m},\mu, p)  = 0  \;,
\end{equation}
where $\Delta_h = 2$ is the scaling dimension of \eqref{eq:h} which corresponds to 
the engineering dimension since $E_\ph$ is a physical observable.
Using the fact that the $p$-dependence is known exactly,
$h= 2 E_\ph^2 =2( M_\ph^2 + \VEC{p}^2)$, one can rewrite \eqref{eq:1} as 
\begin{equation}
\label{eq:sowas}
( \R{\beta} \pd{\R{g}} - \R{m}(1+ \R{\gamma}) \pd{\R{m}} +  \Delta^{\rm eff}_{E^2}) 
E^2_{\ph}  = 0  \;, \quad \Delta^{\rm eff}_{E^2}  \equiv 2\frac{M_\ph^2}{E_\ph^2} \;.
\end{equation}
The two derivatives in \eqref{eq:sowas} can be substituted 
by the relation \eqref{eq:FHfunM} and $m \pd{m} E^2_\ph =    \R{m} \matel{\ph}{\bar{\R{q}}  \R{q}}{\ph}_c$  
(e.g. eq (17)  in \cite{FHfun}) with slight proliferation  of notation in terms of barred quantities in the last expression. One obtains,
\begin{equation}
\label{eq:TA}
2 M_\ph^2 = \frac{ \R{\beta}}{2 \R{g}} \matel{\ph}{\frac{1}{\R{g}^2} \R{G}^2}{\ph}_c + 
(1+ \R{\gamma})  \R{m} \matel{\ph}{ \bar{\R{q}}  \R{q}}{\ph}_c \;,
\end{equation}
which corresponds to the well-known matrix element of the trace anomaly \cite{EMTtrace} between between a physical state (e.g. \cite{FHfun}).  

We note that the derivation in 
this appendix corresponds to the, almost, backwards derivation of \cite{FHfun} where 
the Feynman-Hellmann relation \eqref{eq:FHfunM} is derived from the trace anomaly. 
Furthermore it is also closely related to the heuristic derivation of the trace anomaly 
using $T_{\phantom{x}\alpha}^{\alpha} \propto \frac{d}{d \mu} {\cal L}(\mu)$ where $\mu$ stands for 
some renormalization scale.
The main reason for presenting the derivation is to clarify how matters work 
out for states with non-zero spatial momenta (i.e. $M^2_\ph \neq E_\ph^2$). The latter necessitate an 
RGE \eqref{eq:1} where  the external momenta are taken into account.

\end{document}